\documentclass[epj,nopacs]{svjour}
\usepackage{graphics}
\usepackage{epsfig}
\usepackage{multirow}
\newcommand{\pt}{p_\mathrm{T}}
\newcommand{\Lc}{\mathrm{\Lambda}_\mathrm{C}}
\newcommand{\D}{\mathrm{D}^{0}}
\begin{document}
\title{STAR inner tracking upgrade - A performance study}
\author{Jan Kapit\'an\inst{1} (for the STAR Collaboration)
% \thanks{\emph{Present address:} Insert the address here if needed}%
}                     % Do not remove
\institute{Nuclear Physics Institute ASCR, Rez/Prague, Czech Republic, \email{kapitan@rcf.rhic.bnl.gov}}
\date{Received: date / Revised version: date}
\abstract{
Anisotropic flow measurements have demonstrated development of partonic collectivity in $200~\mathrm{GeV}$ Au+Au collisions at RHIC. To understand the partonic EOS, thermalization must be addressed. Collective motion of heavy-flavor (c,b) quarks can be used to indicate the degree of thermalization of the light-flavor quarks (u,d,s). Measurement of heavy-flavor quark collectivity requires direct reconstruction of heavy-flavor hadrons in the low $\pt$ region. Measurement of open charm spectra to high $\pt$ can be used to investigate heavy-quark energy loss and medium properties.
The Heavy Flavor Tracker (HFT), a proposed upgrade to the STAR experiment at midrapidity, will measure $v_{2}$ of open-charm hadrons to very low $\pt$ by reconstructing their displaced decay vertices. The innermost part of the HFT is the PIXEL detector (made of two low mass monolithic active pixel sensor layers), which delivers a high precision position measurement close to the collision vertex. The Intermediate Silicon Tracker (IST), a 1-layer strip detector, is essential to improve hit identification in the PIXEL detector when running at full RHIC-II luminosity. 
Using a full GEANT simulation, open charm measurement capabilities of STAR with the HFT will be shown. Its performance in a broad $\pt$ range will be demonstrated on $v_{2}$ ($\pt > 0.5~\mathrm{GeV}/c$) and $R_\mathrm{CP}$ ($\pt < 10~\mathrm{GeV}/c$) measurements of $\D$ meson. Results of reconstruction of $\Lc$ baryon in heavy-ion collisions are presented.
%
%\PACS{
%      {PACS-key}{discribing text of that key}   \and
%      {PACS-key}{discribing text of that key}
%     } % end of PACS codes
} %end of abstract
\maketitle
\section{Introduction}
\label{intro}
Produced by the initial hard scattering, heavy quarks (c,b) are an ideal tool to probe early stages of heavy ion collision~\cite{p5}. They derive their mass from the Higgs field, and therefore are not modified by the surrounding QCD medium (they stay heavy even in the case of chiral symmetry restoration).

Previous studies have identified the development of partonic collectivity in heavy-ion collisions at RHIC~\cite{p5}, but they have not yet demonstrated thermalization of the created matter. 
The study of heavy quark collectivity may allow us to address this issue. Measurement of elliptic flow ($v_{2}$) of open charm hadrons to low transverse momentum ($\pt$) is of particular interest.

Suppression of high $\pt$ hadron production at RHIC~\cite{jaro1,jaro2,jaro3} is commonly thought to arise from partonic energy loss in dense matter due to induced gluon radiation~\cite{jaro4}. Radiative energy loss of heavy quarks is expected to be suppressed (dead cone effect)~\cite{jaro11}.

Most heavy-flavor measurements at RHIC use semileptonic decay modes of heavy-flavor hadrons, but lack of precise kinematical information about the parent hadron makes it difficult to study dynamics at low $\pt$~\cite{po1}. Measurements of nuclear modification factor ($R_\mathrm{AA}$) of heavy-flavor decay electrons at high $\pt$~\cite{jaro,phenix} indicate a significant energy loss of heavy quarks. However, knowledge of the relative contributions of charm and bottom decays to electron spectra is crucial to interpret these results. Electron spectra from open charm hadron decays are also sensitive to relative $\Lc / \mathrm{D}$ meson yield~\cite{lc4}, due to different branching ratios of their inclusive electron decay channels.

In central Au+Au collisions at RHIC, a baryon/meson enhancement has been observed in the intermediate $\pt$ region ($2 < \pt < 6~\mathrm{GeV/}c$)~\cite{lc1,lc2}. 
These results are usually explained by a hadronization mechanism involving collective multi-parton coalescence rather than independent vacuum fragmentation. The  success of the coalescence approach implies deconfinement and possibly thermalization of the light quarks prior to hadronization.

Since $\Lc$ is the lightest charmed baryon, and its mass is not far from that of the $\D$ meson, a similar pattern of baryon/meson enhancement is expected in charm sector~\cite{lc3}. $\Lc / \D$ enhancement is also believed to be a signature of a strongly coupled quark-gluon plasma~\cite{lcqgp}. Therefore it would be very interesting to measure $R_\mathrm{CP}$ of $\Lc$ baryon and compare it to $R_\mathrm{CP}$ of $\D$ mesons.

Direct reconstruction of open charm hadrons is necessary for these measurements.
Given the large combinatorial backgrounds in heavy-ion collisions, topological reconstruction is needed to achieve reasonable signal significance.
The Heavy Flavor Tracker (HFT), a proposed upgrade \cite{po4} to the STAR experiment, will enable measurement of open charm hadrons by reconstructing their displaced decay vertices. 

\section{Heavy Flavor Tracker design}
\label{design}
The HFT detector consists of two subsystems: The PIXEL detector (2 layers) and Intermediate Silicon Tracker (IST, 1 layer). The midrapidity tracking system of STAR further includes the existing Silicon Strip Detector (SSD) and the large Time Projection Chamber (TPC).
Compared to the previous version, the current design of the HFT has been optimised for lower mass and better hit resolving, however it has not been fully simulated yet. In the following, simulation results are presented for the previous design. The parameters of these are displayed in Table~\ref{tab:parameters}.

\begin{table}[htb]
\caption{Hit position resolutions of SSD + HFT layers for the two design versions. IST2 (simulated design) has two layers (A, B) with crossed strips.}
\label{tab:parameters}
\begin{center}
\begin{tabular}{|l||c|c||c|c|}\hline
& \multicolumn{2}{c||}{current design} & \multicolumn{2}{c|}{simulated design} \\ \hline
\multirow{3}{*}{layer}& \multirow{3}{*}{r (cm)} & hit resol. & \multirow{3}{*}{r (cm)} & hit resol. \\
& & ($r-\phi \times z$) & &($r-\phi \times z$)  \\ 
& & ($\mu\mathrm{m} \times \mu\mathrm{m}$) & & ($\mu\mathrm{m} \times \mu\mathrm{m}$) \\ 
\hline \hline
SSD & 23 & 30 $\times$ 699 & 23 & 30 $\times$ 699 \\ \hline
IST2-B & - & - & 17 & 17 $\times$ 12000 \\ \hline
IST2-A & - & - & 17 & 12000 $\times$ 17 \\ \hline
IST1   & 14 & 115 $\times$ 2900 & 12 & 17 $\times$ 6000 \\ \hline
PIXEL2 & 8 & 5 $\times$ 5 & 7 & 9 $\times$ 9 \\ \hline
PIXEL1 & 2.5 & 5 $\times$ 5 & 2.5 & 9 $\times$ 9 \\ \hline
\end{tabular}
\end{center}

\end{table}

\begin{figure}[htb]
\centering
\includegraphics[width=0.47\textwidth]{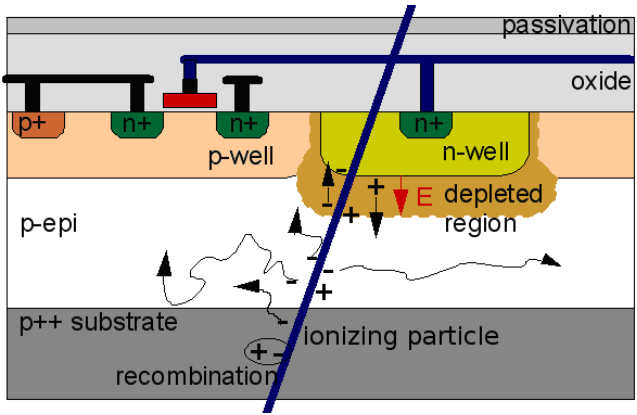}
\caption{MAPS principle of operation: electrons created in the epitaxial layer thermally diffuse towards low potential n-well region. A small contribution to the total signal also exists from electrons created in the p++ substrate.}
\label{fig:maps}
\end{figure}

The PIXEL detector is made of low-mass monolithic active pixel sensors (MAPS, see Fig.~\ref{fig:maps}) and enables high precision measurement close to the primary collision vertex, featuring 18 $\mu$m pixel pitch and a thickness of only $0.28 \%~X_{0}$ per layer. It's fabricated by CMOS technology. 

\begin{figure}[htb]
\centering
\includegraphics[width=0.47\textwidth]{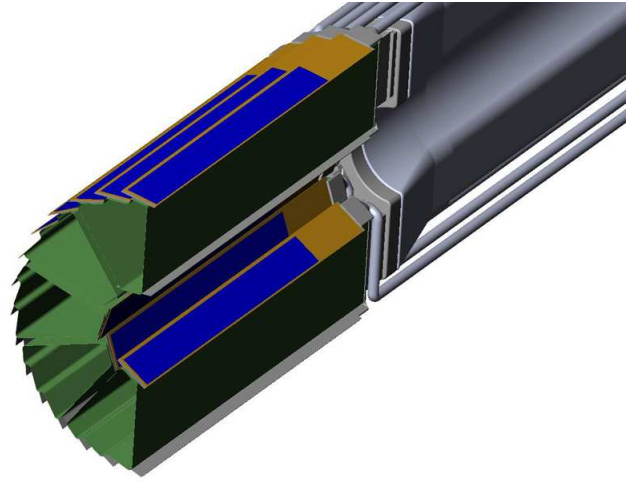}
\caption{One half of the PIXEL detector with its support structure. It will be mounted on one side, where also all the cables and cooling air comes from.}
\label{fig:detector}
\end{figure}

The detector configuration is shown in Fig.~\ref{fig:detector}. Inner layer sensors facing the beam pipe will improve STAR's di-lepton capability by rejecting $\mathrm{e}^{+}\mathrm{e}^{-}$ pairs from photon conversions (beam pipe thickness being only $\approx 0.15 \%~X_{0}$).

As there is a very large number of pixels, the readout of the PIXEL detector is $\approx 200~\mu\mathrm{s}$. Given the high luminosity projected for the future RHIC-II upgrade ($50 \cdot 10^{26} \mathrm{cm}^{-2} \mathrm{s}^{-1}$), the PIXEL detector will integrate over $\approx$ 10 minimum bias collisions. 
The IST detector is essential to improve hit identification at PIXEL2 layer in this pile-up environment. The IST detector consists of 1 layer of fast single-sided strip sensors.

\section{Simulation results}
\label{results}
HIJING central Au+Au events at $\sqrt{s_\mathrm{NN}} = 200~\mathrm{GeV}$ with added $\D$ and $\Lc$ were filtered through GEANT and STAR detector response simulators. To assess the impact of pile-up, pseudo-random hits were added to PIXEL1 and PIXEL2 layers, corresponding to a minimum bias (MB) collision rate of 80 kHz and a primary vertex diamond size $\sigma_{PV\_Z} = 15~\mathrm{cm}$. This is upper estimate of pile-up, in fact current RHIC-II projections show smaller luminosity (corresponding to MB collision rate $\approx 50~\mathrm{kHz}$) and bigger $\sigma_{PV\_Z}$.

The decay modes used for reconstruction were $\D \rightarrow \mathrm{K}^{-}+\pi^{+}$ (B.R. 3.8\%) and $\Lc \rightarrow \mathrm{K}^{-}+\pi^{+}+\mathrm{p}$ (B.R. 5.0\%).
A possibility of using a $\Lc$ decay through a resonant intermediate $\mathrm{\Lambda}(1520)$ state has been investigated, and a significant improvement of S/B ratio could be expected. However, a full simulation hasn't been performed, and results shown hereafter are for the non­resonant decay.

\begin{figure}[htb]
\centering
\includegraphics[width=0.47\textwidth]{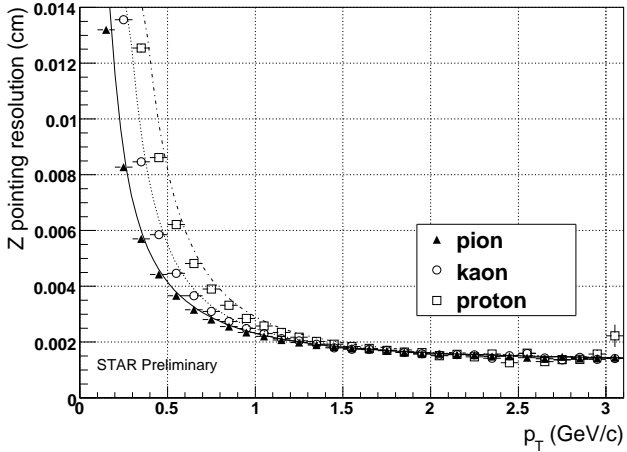}
\caption{Track impact parameter resolution. Symbols: full detector simulation, lines: calculation for PIXEL detector only~\cite{po4}.}
\label{fig:pointing}
\end{figure}

As already mentioned in Section~\ref{design}, the track impact parameter (pointing) resolution is mainly delivered by the PIXEL detector, which is consistent with results from full simulation shown in Fig.~\ref{fig:pointing}.
For particle identification (PID) of daughter particles, $\mathrm{K} - \pi$ separation for $\pt < 1.6~\mathrm{GeV/}c$ and $(\mathrm{K}+\pi) - \mathrm{p}$ separation for $\pt < 3.0~\mathrm{GeV/}c$ is expected using the MRPC TOF detector~\cite{tof}, with an efficiency $\approx 90\%$.

\begin{figure}
\centering
\includegraphics[width=0.47\textwidth]{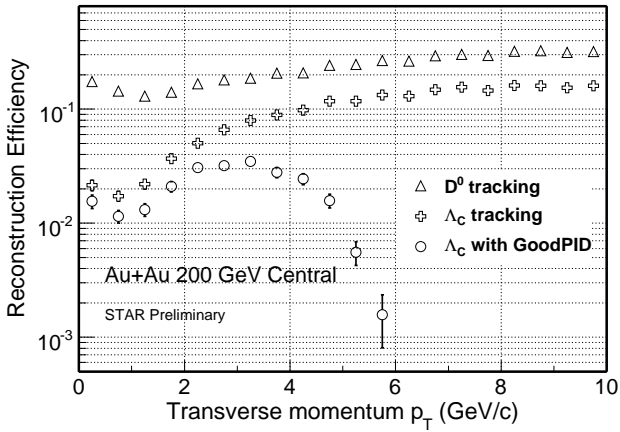}
\caption{Efficiency of reconstructing daughter tracks of $\D$ and $\Lc$ in $|\eta| < 1$. Note that no cuts to isolate signal from combinatorial background have been applied here.}
\label{fig:eff}
\end{figure}

Reconstruction efficiencies for $\D$ and $\Lc$ are shown in Fig.~\ref{fig:eff}. The daughter tracks are required to be well reconstructed in the TPC and have correctly associated hits in both PIXEL layers. Efficiency for $\Lc$ is smaller due to the fact, that there are three daughter tracks with, on average, lower $\pt$ than in the case of $\D$.

For $\Lc$ reconstruction, primary track combinatorial background is huge due to its very short decay length ($c\tau = 59.9~\mu\mathrm{m}$) and three-body decay. To reduce this background, PID information of daughter tracks is required for $\pt < 5~\mathrm{GeV/}c$, limiting the acceptance. It's possible to reconstruct $\Lc$ at higher $\pt$, but this hasn't been studied in detail yet.

Given the overall uncertainty of $c\bar{c}$ cross section at RHIC, $\D$ $dN/dy = 0.002$ per binary collision has been used for signal estimates, which is half of the value measured by the STAR Collaboration \cite{charm}. $\Lc / \D$ ratio $0.2$ was assumed for the case of no enhancement \cite{lc3}. The shape of $\pt$ spectra for $\D$ and $\Lc$ in central Au+Au collisions was assumed to be a power-law, with $\langle \pt \rangle = 1.0~\mathrm{GeV/}c$ and $n=11$.  

To reduce the large combinatorial background, topological cuts have been applied: only tracks with Distance of Closest Approach to event Primary Vertex $DCA_{PV} > DCA_{PV,cut}$ have been used, where the cut is in the range $40 - 80~\mu\mathrm{m}$ depending on transverse momentum of reconstructed $\D$ ($\Lc$). The $DCA$ of daughter tracks to the decay vertex was required to be less than $2 \sigma$ and $\D$ ($\Lc$) momentum was required to point back to the event primary vertex. Finally, a 2 (3) particle invariant mass cut has been applied.

\begin{figure}
\centering
\includegraphics[width=0.47\textwidth]{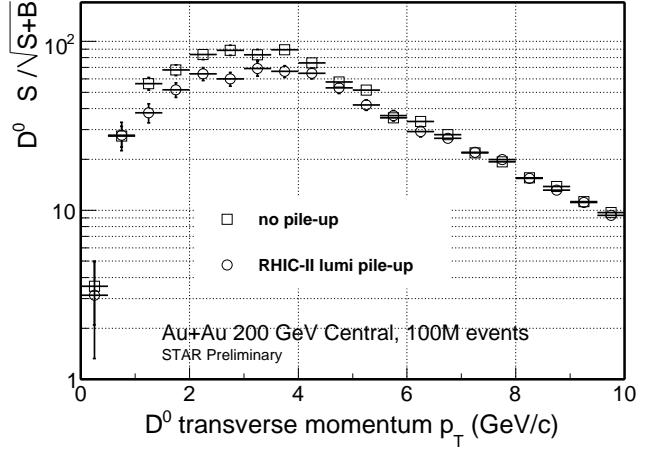}
\caption{$\D$ signal significance in central Au+Au events. The effect of pile-up is visible in low and medium $\pt$ region.}
\label{fig:ss}
\end{figure}

Signal significance of $\D+\overline{\mathrm{D}^{0}}$ from 100M central Au+Au collisions is shown in Fig.~\ref{fig:ss}. 
To estimate background for $\D$ ($\Lc$) at different centralities, $(N_{part})^{2}$ ($(N_{part})^{3}$) scaling was used. For the signal, the same $R_\mathrm{CP}$ was assumed as measured by STAR for charged hadrons~\cite{jaro2}. This gives a factor $4$ higher yield at high $\pt$ in peripheral (60-80\%) collisions than given by $N_{bin}$ scaling. Ratio of $N_{bin}$ between central (0-10\%) and peripheral (60-80\%) collision is $\approx 44$.     

For measurements of $\D$ $v_{2}$ and $R_\mathrm{CP}$, statistics of 500M minimum bias Au+Au events was assumed. This is the amount of data the STAR detector can take in about 1 month of running (with DAQ rate 500~Hz and 40\% beam up time).

\begin{figure}[htb]
\centering
\includegraphics[width=0.47\textwidth]{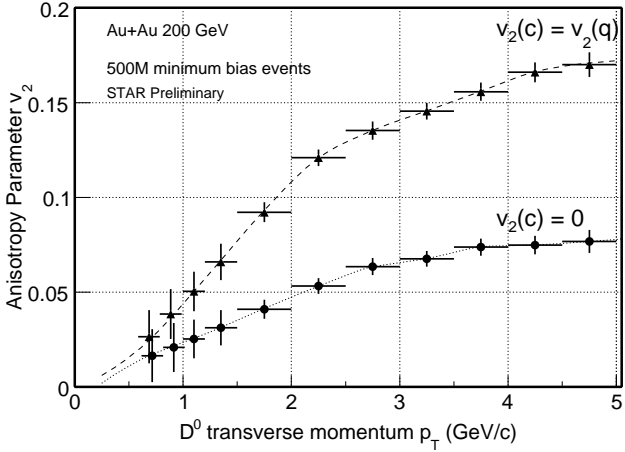}
\caption{Two scenarios for $\D$ elliptic flow (charm quark flow zero or equal to the flow of light quarks) and estimated statistical errors.}
\label{fig:v2}
\end{figure}

Estimated statistical errors on $v_{2}$ (for $\D+\overline{\mathrm{D^{0}}}$, $|\eta|<1$) are shown in Fig.~\ref{fig:v2}. For $\pt > 1.0~\mathrm{GeV/}c$, the HFT will be able to distinguish between extreme elliptic flow scenarios in the coalescence picture in the first year of operation.

\begin{figure}[htb]
\centering
\includegraphics[width=0.47\textwidth]{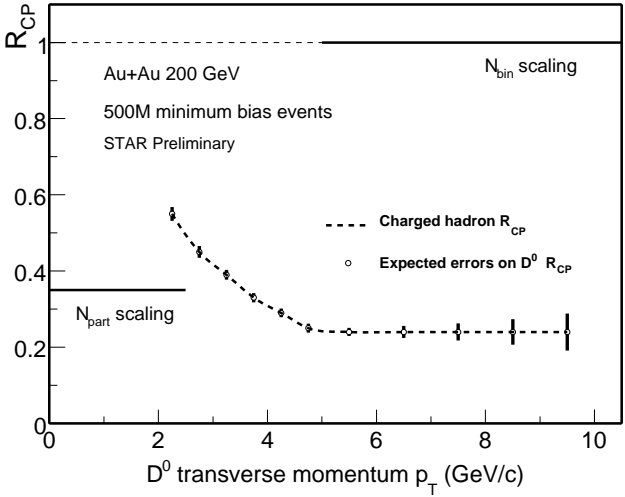}
\caption{Estimated statistical errors for $R_\mathrm{CP}$ measurement of $\D$ meson, in 500M minimum bias Au+Au events.}
\label{fig:Rcp}
\end{figure}

500M minimum bias events include 62.5M central and 125M peripheral events for $R_\mathrm{CP}$ measurement. $\D$ signal significance (100M central events) in $9 < \pt < 10~\mathrm{GeV}/c$ bin is about $15 \sigma$ (consistent with Fig.~\ref{fig:ss}). The background is negligible here, so signal significance in 100M peripheral events will be $\sqrt(44/4) \approx 3.3$ times smaller. In a similar way, the calculation was done for lower $\pt$ (taking into account the background here), and the estimated statistical errors on $\D$ $R_\mathrm{CP}$ measurement are shown in Fig.~\ref{fig:Rcp}. At highest $\pt$, the relative error is $20\%$, which is a good precision for this region.  

\begin{figure}[htb]
\centering
\includegraphics[width=0.47\textwidth]{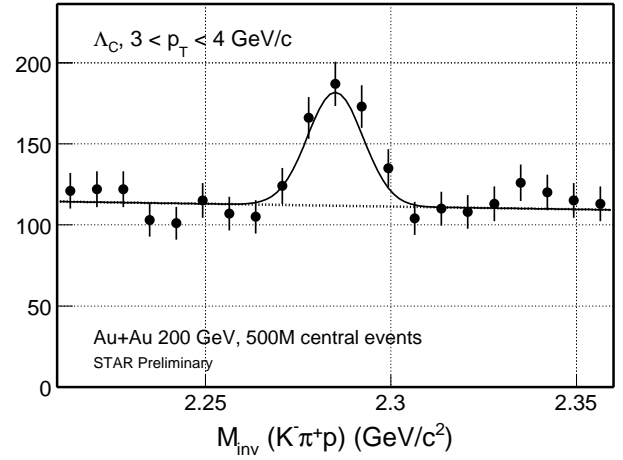}
\caption{Expected invariant mass distribution of $\Lc$, $3 < \pt < 4~\mathrm{GeV/}c$, for 500M central events.}
\label{fig:minv}
\end{figure}

For measurements of $\Lc$, statistics of 2B minimum bias and 250M central triggered events was assumed, giving 500M central and 500M peripheral events used for $R_\mathrm{CP}$. 
$\Lc$ reconstruction cuts were optimized in $\pt$ range $2-5~\mathrm{GeV/}c$, estimated mass peak for the middle $\pt$ bin is shown in Fig.~\ref{fig:minv}. Estimated signal significance for this $\pt$ bin in 500M central Au+Au collisions is $8 \sigma$.

To estimate errors of $\Lc / \D$ measurement, two scenarios have been used: 1. no enhancement, ratio equal to $0.2$ and flat in $\pt$, 2. the same enhancement as $\mathrm{\Lambda} / \mathrm{K}^{0}_{S}$~\cite{lc2}. $\D$ has much larger yield and $c\tau$ than $\Lc$, statistical errors coming from its measurement are therefore negligible. $\Lc / \D$ ratio in peripheral collisions is assumed flat (in rough agreement with $\mathrm{\Lambda} / \mathrm{K}^{0}_{S}$~\cite{lc2}), with statistical errors given by rescaling signal and background from central to peripheral collisions, as describer above. 

\begin{figure}
\centering
\includegraphics[width=0.47\textwidth]{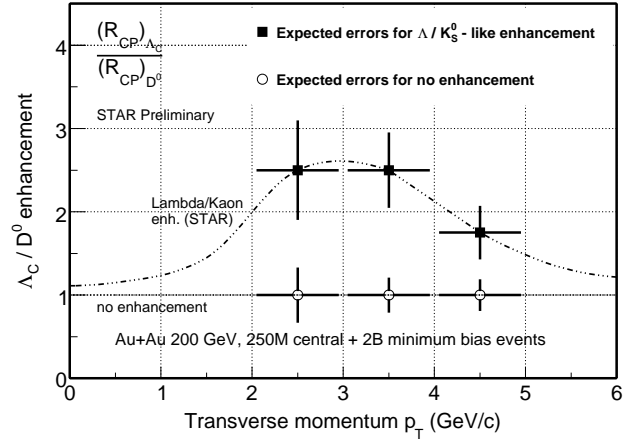}
\caption{Expected errors on $\Lc / \D$ measurement, assuming 500M central and 500M peripheral collisions. Except for the lowest $\pt$ bin, the errors are dominated by measurement of $\Lc$ in peripheral collisions.}
\label{fig:lcd0}
\end{figure}

Statistical errors on $R_\mathrm{CP} (\Lc) / R_\mathrm{CP} (\D)$ are shown in Fig.~\ref{fig:lcd0}, the two extreme cases can be well distinguished. Thus, baryon/meson ratio in charm sector will be measured with good precision, for the first time in heavy ion collisions.

\section{Conclusions}
\label{summary}

The HFT detector will measure open charm hadrons over a broad $\pt$ range, enabling precision study of charm collectivity, energy loss and baryon/meson ratios. These are important ingredients for a systematic study of the dense medium created in heavy-ion collisions at RHIC.

This will be achieved by using low mass MAPS sensors (PIXEL) together with a fast strip detector (IST), delivering high efficiency and ultimate pointing resolution at low $\pt$, even in the high luminosity environment of RHIC-II. Improved reconstruction efficiency is expected due to improved detector design and due to possibly lower pile­up hit densities in RHIC-II luminosity, to be fully simulated in the future.

\section*{Acknowledgement}
\label{acknowledgement}

This work was supported in part by the IRP AV0Z10480505, 
by GACR grant 202/07/0079 and by grant LC07048 of the Ministry 
of Education of the Czech Republic.


\begin{thebibliography}{}

\bibitem{p5}
J.~Adams {\it et al.}, 
Nucl.\ Phys.\ A \textbf{757}, (2005) 102.

\bibitem{jaro1}
C.~Adler {\it et al.},
Phys.\ Rev.\ Lett. \textbf{89}, (2002) 202301.

\bibitem{jaro2}
J.~Adams {\it et al.},
Phys.\ Rev.\ Lett. \textbf{91}, (2003) 172302.

\bibitem{jaro3}
J.~Adams {\it et al.},
Phys.\ Rev.\ Lett. \textbf{91}, (2003) 072304.

\bibitem{jaro4}
R.~Baier {\it et al.},
Ann.\ Rev.\ Nucl.\ Part.\ Sci. \textbf{50}, (2002) 37;
M.~Gyulassy {\it et al.}, arXiv: nucl-th/0302077.
 
\bibitem{jaro11}
Y.~L.~Dokshitzer and D.~E.~Kharzeev, 
Phys.\ Lett.\ B \textbf{519}, (2001) 199.

\bibitem{po1}
S.~Batsouli, S.~Kelly, M.~Gyulassy and J.~L.~Nagle,
Phys.\ Lett.\ B \textbf{557}, (2003) 26.

\bibitem{jaro}
B.~I.~Abelev {\it et al.},
Phys.\ Rev.\ Lett. \textbf{98}, (2007) 192301.

\bibitem{phenix}
A.~Adare {\it et al.},
Phys.\ Rev.\ Lett. \textbf{98}, (2007) 172301.

\bibitem{lc4}
P.~R.~Sorensen and X.~Dong,
Phys.~Rev.~C \textbf{74}, (2006) 024902.

\bibitem{lc1}
B.~I.~Abelev {\it et al.},
Phys.\ Rev.\ Lett. \textbf{97}, (2006) 152301;
B.~I.~Abelev {\it et al.},
Phys.\ Lett.\ B \textbf{655}, (2007) 104.

\bibitem{lc2}
J.~Adams {\it et. al.},
arXiv: nucl-ex/0601042.
%M.~A.~C.~Lamont {\it et. al.}, 
%in {\em Proceedings of SQM'06 Conference, LA, 2006} (J.~Phys.~G \textbf{32}, (2006)), S105.

\bibitem{lc3}
A.~Andronic {\it et. al.}, 
Phys.~Lett.~B \textbf{571}, (2003) 36.

\bibitem{lcqgp}
S.~H.~Lee {\it et. al.}, 
arXiv: 0709.3637v3 [nucl-th].

\bibitem{po4}
E.~Anderssen {\it et. al.},
{\em A Heavy Flavor Tracker for STAR}, available at http://rnc.lbl.gov/hft/docs/hft\_final\_submission\_version.pdf.

%\bibitem{rhicII}
%W.~Fischer {\it et. al.},
%{\em RHIC Collider Projections (FY2009 - FY2013)}, available at http://www.agsrhichome.bnl.gov/RHIC/Runs/RhicProjections.pdf.

\bibitem{tof}
O.~Barannikova {\it et. al.},
Nucl.~Instrum.~Meth.~A \textbf{558}, (2006) 419-429.

\bibitem{charm}
B.~I.~Abelev {\it et. al.},
arXiv: 0805.0364v2 [nucl-ex].

\end{thebibliography}
\end{document}